\documentclass[pra,aps,superscriptaddress,twoside,twocolumn,notitlepage]{revtex4-1}

\usepackage{graphicx,epic,eepic,epsfig,amsmath,latexsym,amssymb,verbatim,color,bm}

\newcommand{\be}{\begin{eqnarray}}
\newcommand{\ee}{\end{eqnarray}}

\usepackage{theorem}
\newtheorem{definition}{Definition}

\newtheorem{theorem}[definition]{Theorem}

\def\squareforqed{\hbox{\rlap{$\sqcap$}$\sqcup$}}
\def\qed{\ifmmode\squareforqed\else{\unskip\nobreak\hfil
\penalty50\hskip1em\null\nobreak\hfil\squareforqed
\parfillskip=0pt\finalhyphendemerits=0\endgraf}\fi}
\def\endenv{\ifmmode\;\else{\unskip\nobreak\hfil
\penalty50\hskip1em\null\nobreak\hfil\;
\parfillskip=0pt\finalhyphendemerits=0\endgraf}\fi}

\mathchardef\ordinarycolon\mathcode`\:
\mathcode`\:=\string"8000
\def\vcentcolon{\mathrel{\mathop\ordinarycolon}}
\begingroup \catcode`\:=\active
  \lowercase{\endgroup
  \let :\vcentcolon
  }

\newcommand{\nc}{\newcommand}
\nc{\rnc}{\renewcommand}
\nc{\beq}{\begin{equation}}
\nc{\eeq}{{\end{equation}}}
\nc{\beqa}{\begin{eqnarray}}
\nc{\eeqa}{\end{eqnarray}}
\nc{\lbar}[1]{\overline{#1}}
\nc{\bra}[1]{\langle#1|}
\nc{\ket}[1]{|#1\rangle}
\nc{\ketbra}[2]{|#1\rangle\!\langle#2|}
\nc{\braket}[2]{\langle#1|#2\rangle}
\nc{\proj}[1]{| #1\rangle\!\langle #1 |}
\nc{\avg}[1]{\langle#1\rangle}
\nc{\Rank}{\operatorname{Rank}}
\nc{\smfrac}[2]{\mbox{$\frac{#1}{#2}$}}
\nc{\tr}{\operatorname{Tr}}
\nc{\ox}{\otimes}
\nc{\dg}{\dagger}
\nc{\dn}{\downarrow}
\nc{\cA}{{\cal A}}
\nc{\cB}{{\cal B}}
\nc{\cC}{{\cal C}}
\nc{\cD}{{\cal D}}
\nc{\cE}{{\cal E}}
\nc{\cF}{{\cal F}}
\nc{\cG}{{\cal G}}
\nc{\cH}{{\cal H}}
\nc{\cI}{{\cal I}}
\nc{\cJ}{{\cal J}}
\nc{\cK}{{\cal K}}
\nc{\cL}{{\cal L}}
\nc{\cM}{{\cal M}}
\nc{\cN}{{\cal N}}
\nc{\cO}{{\cal O}}
\nc{\cP}{{\cal P}}
\nc{\cR}{{\cal R}}
\nc{\cS}{{\cal S}}
\nc{\cT}{{\cal T}}
\nc{\cX}{{\cal X}}
\nc{\cZ}{{\cal Z}}
\nc{\csupp}{{\operatorname{csupp}}}
\nc{\qsupp}{{\operatorname{qsupp}}}
\nc{\var}{{\operatorname{var}}}
\nc{\rar}{\rightarrow}
\nc{\lrar}{\longrightarrow}
\nc{\polylog}{{\operatorname{polylog}}}
\nc{\1}{{\openone}}
\nc{\wt}{{\operatorname{wt}}}
\nc{\av}[1]{{\left\langle {#1} \right\rangle}}

\nc{\RR}{{{\mathbb R}}}
\nc{\CC}{{{\mathbb C}}}
\nc{\FF}{{{\mathbb F}}}
\nc{\NN}{{{\mathbb N}}}
\nc{\ZZ}{{{\mathbb Z}}}
\nc{\PP}{{{\mathbb P}}}
\nc{\QQ}{{{\mathbb Q}}}
\nc{\UU}{{{\mathbb U}}}
\nc{\EE}{{{\mathbb E}}}
\nc{\id}{{\operatorname{id}}}

\begin{document}

\title{``Hyperbits'': the information quasiparticles}

\author{Marcin Paw\l{}owski}
\affiliation{Department of Mathematics, University of Bristol, Bristol BS8 1TW, U.K.}
\email{maymp@bris.ac.uk}

\author{Andreas Winter}
\affiliation{Department of Mathematics, University of Bristol, Bristol BS8 1TW, U.K.}
\affiliation{Centre for Quantum Technologies, National University of Singapore, 2 Science Drive 3, Singapore 117542}
\email{a.j.winter@bris.ac.uk}

\date{7 October 2011}

\begin{abstract}
Information theory has its particles, bits and qubits, just as physics has electrons and photons. However, in physics we have a special category of objects with no clear counterparts in information theory: quasiparticles. They are introduced to simplify complex emergent phenomena making otherwise very difficult calculations possible and providing additional insight into the inner workings of the system. We show that we can adopt a similar approach in information theory. We introduce the hyperbits, the first information quasiparticles which we prove to be a resource equivalent to entanglement and classical communication, and give examples how they can be used to simplify calculations and get more insight into communication protocols.
\end{abstract}

\maketitle

\section{Introduction: Hyperbits}

Recently much work has been done towards the end of deriving quantum theory from information-theoretic principles (see e.g. \cite{Hardy:simple,Brukner:simple,Masanes:simple}). Although the choice of axioms and the details of the derivations are different, all these papers have one thing in common. The authors are able to show that the structure underlying quantum theory should be based on Hilbert space. Explicitly, they show that states and measurements can be represented by vectors from this space and the probabilities of experimental outcomes expressed are by the familiar formula involving the scalar product. For the simplest system this is equivalent to the derivation of the Bloch sphere. Projective measurements are represented by normalized vectors while the vectors for the states can have any length smaller or equal $1$. The expectation value of the measurement is then given by the scalar product of these two vectors.

The part of the derivation of the quantum theory where it is shown that states and measurements are vectors in some Euclidean ball, is relatively easy, natural and straightforward in all the papers dealing with this problem. What is not so easy is showing that the dimension of the Bloch sphere is $3$. Of course, in all the papers mentioned the authors are somehow able to overcome this difficulty and find reasons for this particular dimension, but the derivation at this point always looses some of its compelling elegance. It is then only natural that toy theories with higher dimensional Bloch spheres as state spaces have also been studied  \cite{Karol,U1,U2,LIC}. In such theories the formula for the probabilities of experimental outcomes stays the same. To wit, with the state given by a vector $\vec{v}$ of length $| \vec{v} | \leq 1$ in some $d$-dimensional Euclidean space, a measurement with outcomes $X=\pm 1$ described by the pair $\pm\vec{w}$ with a unit vector $\vec{w}$, the measurement outcome is a random variable with expectation
\[
  \langle X \rangle = \braket{w}{v}.
\]
In other words, compared to usual quantum mechanical Bloch sphere,
the only thing that changes is the allowed dimensionality of the vectors describing the state and the measurement. A system that can be in any state described by vectors of any finite dimension we call a \emph{hyperbit}. The choice of the name follows from the fact that its state space is a hyperball.

While these objects seems a priori of limited theoretical interest, we find here that, remarkably, hyperbits do exist! However, their existence is only an effective one. They resemble the quasiparticles from solid state physics, such as holes or phonons. They are introduced to make calculations, otherwise complicated, simple and at the same time give intuitive explanations of the physical phenomena observed.

Just as in the case of those standard quasiparticles, one needs to find an experimental setting where they become useful and prove that a single quasiparticle can be substituted for a complex emergent phenomenon. For the hyperbits the setting is the very general case of a communication protocol of the following type: There are two parties, called sender and receiver, each of which receives an \emph{input}, given as a binary string or a number, then they are allowed to make use of resources such as shared entangled states, and are allowed the sending of a single bit from sender to receiver; at the end the receiver outputs an answer consisting again of a single bit.
We do not make any assumption on the size of the input of neither the sender nor the receiver. In this setting we are able to show that sharing arbitrary entangled states and one bit of classical communication from the sender to the receiver is equivalent to the communication of one hyperbit in the same direction.

This result may be interesting in itself from a conceptual point of view as it shows that there exists indeed something like an information quasiparticle. But the quasiparticle's {\it raison d'\^{e}tre} is simplifying some calculations so there would be a little point in introducing the hyperbits if we could not provide at least one example where they do just that. Fortunately, we have more than one.

The paper is structured as follows. First, we prove that for tasks where Bob gives binary answers, sending one hyperbit from Alice to Bob is equivalent to sharing any amount of entanglement and sending (also from Alice to Bob) one classical bit. This theorem is the main result of this paper; note that the scenario where the communication of the hyperbit can be substituted appears in the studies of communication complexity \cite{BCvD,BZ}, oblivious transfer \cite{OT}, parity oblivious multiplexing \cite{mpx} or random access codes \cite{RAC,EARAC}. Hyperbits can be used as a tool in all these cases. We back this claim up by providing two examples where using hyperbits allows us to prove a very general identity and a strengthened version of information causality \cite{IC}. We conclude by discussing possible extensions of the main theorem that would make it even more useful. Throughout the paper we assume that the two possible values of a bit are $\pm 1$ instead of the usual $0/1$.

\section{Equivalence with entanglement and classical communication}

In the formulation of our main theorem below we consider \emph{tasks} for
two cooperating players, Alice and Bob. These involve inputs $\vec{a}$ to
Alice and $\vec{b}$ to Bob (for instance bit strings), and require
Bob to output an answer $B$. To do this, they may invoke certain
resources, such as shared entanglement or communication, and their
success will be measured by a referee looking at $\vec{a}$, $\vec{b}$
and $B$ (and possibly outputs of Alice).

\begin{theorem}
  For tasks where Bob gives binary answers, sending one hyperbit from Alice
  to Bob is equivalent to sharing any amount of entanglement and sending
  (also from Alice to Bob) one classical bit.
\end{theorem}

We give a formal proof of this theorem in Appendix A, and here only an outline of it,
with some comments on the interesting points.

In \cite[Thm. 2.1]{Tsirelson:vectors}, Tsirelson uses the properties of Clifford algebra to show that there is a one-to-one correspondence between the set of Hermitian operators and real vectors. More precisely, the equivalence concerns any two sets of Hermitian operators $\{\hat{A}_k\}$ and $\{\hat{B}_m\}$ such that for all $k$ and $m$,
\begin{equation}
  -\1\leq \hat{A}_k\leq \1, \quad -\1\leq \hat{B}_m\leq \1, \quad [\hat{A}_k,\hat{B}_m]=0.
\end{equation}
It states that for any bipartite state $\rho$ for each of these operators there is a corresponding
real vector $\vec{x}_k$ (for $\hat{A}_k$) or $\vec{y}_m$ (for $\hat{B}_m$) such that $|\vec{x}_k|\leq 1$, $|\vec{y}_m|\leq 1$ and
\begin{equation}
  \label{tsi2}
  \langle \vec{x}_k,\vec{y}_m\rangle = \tr(\hat{A}_k\otimes\hat{B}_m\rho).
\end{equation}
This theorem also says that for any two sets of vectors $\{\vec{x}_k\}$ and $\{\vec{y}_m\}$ such that $|\vec{x}_k|\leq 1$ and $|\vec{y}_m|\leq 1$ there exists a state and two sets of hermitian operators such that (\ref{tsi2}) holds.

If the operators considered are projectors, one can then think about the application of the operator $\hat{A}_k$ as the preparation of the hyperbit in the state $\vec{x}_k$. The application of the operator $\hat{B}_m$ is then the measurement of the same hyperbit using the vector $\vec{y}_m$. The only problem is that Alice and Bob cannot use the projective operators of their choice on the state they share. They can only choose the sets of (at least) two different projectors such that one of them will be randomly applied. In other words, they can choose the observables not the specific projectors. But Alice still knows which projector has been applied from the outcome of her measurement so she knows which of the two possible hyperbits she has prepared. If she sends her outcome to Bob then he can adopt his strategy to compensate for the randomness of the state preparation. The proof in the Appendix A shows that this adaptation can always be done.

\medskip
One important property of simulating the entanglement and communication based protocol with the hyperbits is that one has to consider only a particular kind of the postprocessing of the data in the hyperbit case. After the communication of the hyperbit it has to be measured and then some function has to be applied to the measurement outcome. The explicit construction of the simulation protocol presented in Appendix A shows that this function is linear in the outcome. This can significantly simplify calculations where the optimizations over all possible protocols are made.

In the following two sections we show two applications of this theorem.

\section{An identity limiting guessing probabilities}

Let us consider a very general protocol, in the course of which
Alice sends one hyperbit to Bob. From the
previous section we know that it is equivalent to sharing entanglement and
communicating one classical bit. Obviously it is also strictly stronger than
sending a single qubit. The state of the hyperbit encodes information about
Alice's input $\vec{a}$. Bob, in measuring the hyperbit, makes queries about $\vec{a}$.
We assume nothing about the encoding of Alice; the only things we assume are:
\begin{enumerate}
  \item Bob's are yes/no queries: in other words, his outcome is binary.
    Let us label the inputs of Alice by numbers $j=1,..,2^n$ with $p_j$ being the
    probability of Alice getting input $j$. If the number of her possible inputs is
    not a power of 2, we can artificially make it so by adding some more inputs with
    probabilities of occurrence equal 0. We can then construct a matrix $F$ encoding
    all the queries and answers: The correct answer to query $I_i$ when Alice got
    input $j$ is encoded as $f_{i,j} = \pm 1$.
  \item Bob's queries form a complete {\it pairwise unbiased} set. Pairwise unbiasedness
    means that for all $i\neq i'$ the correct answer to queries $I_i$ and $I_{i'}$ i
    s the same for exactly half of the possible inputs of Alice. It does not mean that
    all the bits of Alice's input that Bob can query have to be independent.
    For example if the input of Alice consists of two independent bits $a=\{a_0,a_1\}$,
    Bob can query $a_0$, $a_1$ or $a_0 \oplus a_1$. Completeness means that the
    number of queries is $2^n$. This makes $F$ a square matrix.
  \item Bob's strategy is optimal for the encoding used by Alice: he maximizes his
    probability of guessing each query correctly, under the constraints posed by
    the hyperbit formalism.
  \item The probability distribution of Alice's input is not pathological. One might imagine a case where the most of $p_j$ are 0 or very close to it. This can effectively lead to the case where the correct answers for some queries are always or almost always the same. Bob does not have to get any communication from Alice to give the correct answer. Therefore, we assume that for all the queries the optimal strategy of Bob is measuring the hyperbit, not discarding it and producing perviously prepared value. This does not reduce the generality of the result as Bob can also ask the other queries but they will simply not enter the formulas and calculating the guessing probabilities for them is trivial.
\end{enumerate}

We will use
\begin{equation}
P(I_i=\beta)=\sum_j \frac{|f_{i,j}+\beta|}{2}p_j
\end{equation}
to denote the probability that the correct answer to Bob's query $I_i$ is $\beta = \pm 1$.
If $B_i$ is the answer that Bob gets when he queries $I_i$, then his probability
of the success is $P(B_i=I_i)$ and its bias
\begin{equation}
E_i=2P(B_i=I_i)-1.
\end{equation}

\begin{theorem}
For prior probabilities $p_j$ and hyperbits $\vec{h}_j$ encoding
the messages $j$, the biases $E_i$ of guessing some pairwise
unbiased properties $I_i$ satisfy the identity
\begin{equation} \label{e2}
\sum_{i=1}^{2^n} E_i^2=2^n\sum_{j=1}^{2^n} p_j^2 |\vec{h}_j|^2.
\end{equation}
\end{theorem}
For the proof see Appendix B.

\medskip
This is a very general result, because it is an identity that holds for every distribution of Alice's inputs, every possible encoding and
every complete set of pairwise independent queries. Let us now look at some special cases.

If Alice wants to help Bob her optimal strategy is to always use maximally long hyperbits $|\vec{h'}_j|=1$.
If, furthermore, her inputs have a uniform probability distribution $p_j=2^{-n}$ then
eq.~(\ref{e2}) becomes the neat
\begin{equation}
\sum_{i=1}^{2^n} E_i^2=1.
\end{equation}

Now let us consider the case when $F=H_{2^n}$, in other words it does consist of a row without any $-1$'s. Let's assume it is the first
row. Query $I_1$ corresponding to this row is the only one which gives Bob no information about the input of Alice. If we look at the
vector $\vec{x}_{avg}=\sum_j p_j \vec{h}_j$ we see that it is the average hyperbit that Bob receives. Its length is a measure of what Bob
knows about the message before he receives it and it quantifies, together with the lengths of the individual hyperbits, the optimality of
Alice's encoding. In this case it is meaningful to split the LHS of eq.~(\ref{e2}) into
\begin{equation}
\sum_{i=2}^{2^n} E_i^2+|\vec{x}_{avg}|^2=2^n\sum_{j=1}^{2^n} p_j^2 |\vec{h}_j|^2.
\end{equation}

One might also be interested in the case where the measurements performed by Bob are not optimal
(it may happen that he does not know the particulars of Alice's encoding). Then we obtain
the inequality
\begin{equation}
\sum_{i=1}^{2^n} E_i^2 \leq 2^n\sum_{j=1}^{2^n} p_j^2 |\vec{h}_j|^2,
\end{equation}
instead of (\ref{e2}).

One particular example, where the hyperbits are useful is the study of the security of quantum key distribution against individual attacks. As a simple case study consider the proof presented in \cite{SemiDIQKD}. It is based on the inequality following from corollary 5.2.3 from \cite{konig}
\begin{equation}\label{konig}
P(a_0)+P(a_1)+P(a_0\oplus a_1) \leq \frac{3}{2}\left(1+\frac{1}{\sqrt{3}} \right), \end{equation}
where $P_(a_i)$ are the probabilities of Bob guessing correctly bit $a_i$ when he has received only one bit/qubit of data from Alice. Using Theorem 2 we can obtain a stronger version of the above
\begin{equation}
E(a_0)^2+E(a_1)^2+E(a_0\oplus a_1)^2 \leq 1,
\end{equation}
which leads to weaker requirements on the protocol for the same level of security.

\section{Strengthened information causality}

\emph{Information causality}~\cite{IC} states that if Alice has $N$ independent bits $a_i$ and sends one classical bit to Bob, who can
try to guess the value one of them, then
\begin{equation} \label{ic}
\sum_{i=1}^N I(a_i:b_i)\leq 1.
\end{equation}
$I(a_i:b_i)$ is the mutual information between the bit of Alice and the random variable $b_i$ generated by Bob, which is his best guess
of that bit. In~\cite{IC} we have shown that this property is always true in classical and quantum theory and used it to derive the
Tsirelson bound.

Here we propose a strengthened version of this principle,
proved in Appendix C:

\begin{theorem}
In the scenario described above, inequality~(\ref{ic}) holds even if
the bits $a_i$ of Alice are only pairwise independent.
\end{theorem}

Note also that in~\cite{IC} we were considering the general case of Alice sending
$m$ bits of classical information, though for the derivation of the Tsirelson bound $m=1$ was enough.
Here we only considered the case of $m=1$. In fact one can easily show that for the
communication of $m$ bits $\sum_{i=1}^N I(a_i:b_i)$ can be as large as $2^m-1$,
which seems to be the maximum -- although we are not able to show this yet.

\section{Conclusion}

We have introduced the concept of the information quasiparticle and showed the first example - the hyperbits. We have established two criterions for the introduction of a quasiparticle: the meaningfulness and the usefulness. The first of them requires a specific condition in which it makes sense to substitute the quasiparticle for some more complex emergent phenomenon. This allows for more insight in the workings of the phenomenon studied. The usefulness criterion is more practical. It requires that the introduction of the quasiparticle does help in some calculations. We have shown that the hyperbits satisfy both of these criterions. The more complex case which they can be substituted for is the communication of a single classical bit augmented with any amount of entanglement. Then we  demonstrated
how this substitution can be used to obtain powerful results.
We have provided two examples of such results:
The first is a very general identity which can be used in a wide variety of problems,
including one possible application in cryptography.
The second example is the strengthened version of information causality. Its consequences are of a more foundational type and beyond the scope of this paper. The study of these consequences is one of many direction for future work that the hyperbits open up.
Some of the interesting questions that arise are: Is there a generalization of hyperbits that are equivalent to the scenario of unlimited entanglement plus communication of some fixed finite number of bits? Is there an extension of the generalized information causality bound to communication of more than one bit? Does Theorem 1 also hold if the answers that Bob returns are not binary? Are there any other information quasiparticles?

\acknowledgements
MP is supported by the U.K.~EPSRC and in part
by the Philip Leverhulme Trust and EC integrated
project QESSENCE.
AW is supported by the European Commission, the ERC,
the U.K.~EPSRC, the Royal Society and a Philip Leverhulme Prize.
The Centre for Quantum Technologies is funded by the Singapore
Ministry of Education and the National Research Foundation as part
of the Research Centres of Excellence programme.

\vfill\pagebreak

\appendix

\section{Appendix A: Proof of Theorem 1}

First we show that if Bob gives binary answers then any protocol that uses any amount of entanglement and one bit of classical
communication can be simulated by sending one hyperbit. By simulation we mean another protocol which, for any input $\vec{a}$ of Alice
and $\vec{b}$ of Bob yields the same expectation value of Bob's answer $B$ (for binary outcomes the expectation value carries the whole
information about the probability distribution). Since the communication in the protocol to be simulated is one bit, in the most general
protocol, the behavior of Alice can be considered as just a two outcome measurement of her part of the entangled system and sending her
outcome $A$ to Bob. The choice of Alice's measurement operator depends on her part of the input. Since Bob's answer is also binary his
most general strategy is also to make a two outcome measurement of the system. The choice of Bob's measurement can depend not only on his
input but also on Alice's message.

If Alice and Bob share a uniformly random bit $C$ (in this paper we assume that shared randomness is free),
then the parties can modify their protocol as follows. Alice can make the same
measurement as in the unmodified version and send to Bob the message $A'=CA$. Bob then can get $A$ by again multiplying $A'$ by $C$ and
proceed exactly as he would before the modification. Obviously such protocol gives exactly the same expectation value of $B$. The
difference is that, regardless of what it was before the modification, the expectation value of the message is 0. Such post-processing (in
Alice's case) or pre-processing (in Bob's) can be incorporated into the measurement operators. Therefore, without loss of generality
we may consider only protocols where the expectation value of Alice's outcome $A$ is 0.

Since we do not bound the dimension of the quantum system measured by Alice and Bob, one more thing we can assume without loss of generality is that the measurements they make are projective.

Let $\rho$ be the state that the communicating parties initially share and $\tr_A$ denote partial trace over Alice's subsystem. The measurement that she makes when receiving input $\vec{a}$ is specified by the observable
\begin{equation} \label{hatA}
\hat{A}_{\vec{a}}=\hat{P}_{\vec{a},1}-\hat{P}_{\vec{a},-1},
\end{equation}
where $\hat{P}_{\vec{a},\pm 1}$ are projectors corresponding to the outcomes $\pm 1$.

The expectation value of $B$ given the inputs $\vec{a}$ and $\vec{b}$ and the message $A$ is
\begin{equation}
\langle B(\vec{a},\vec{b},A)\rangle = \tr(\hat{B}_{\vec{b},A}\rho_{\vec{a},A}),
\end{equation}
where $\hat{B}_{\vec{b},A}$ is the observable measured by Bob when his input is $\vec{b}$ and Alice's message $A$.
\begin{equation}
\rho_{\vec{a},A} = \frac{1}{\tr(\hat{P}_{\vec{a},A}\otimes \openone\rho)} \tr_A (\hat{P}_{\vec{a},A}\otimes \openone\rho)
\end{equation}
is Bob's part of the state after Alice's measurement. Our assumption about the expectation value of $A$ being 0 means that for all
$\vec{a}$
\begin{equation}
\tr(\hat{P}_{\vec{a},A}\otimes \openone\rho)=\frac{1}{2}.
\end{equation}
Since
\begin{equation}
\hat{P}_{\vec{a},A}=\frac{\openone +A\hat{A}_{\vec{a}}}{2},
\end{equation}
we get
\begin{equation}\begin{split}
  \langle B(\vec{a},\vec{b},A)\rangle
           &= 2\tr(\hat{P}_{\vec{a},A}\otimes\hat{B}_{\vec{b},A}\rho)                          \\
           &= \tr\Big(\big(\openone +A\hat{A}_{\vec{a}}\big)\otimes \hat{B}_{\vec{b},A}\rho\Big).
\end{split}\end{equation}

Tsirelson ~\cite[Theorem 2.1]{Tsirelson:vectors} considers any two sets of Hermitian operators $\{\hat{A}_k\}$ and $\{\hat{B}_m\}$ such that for all $k$ and $m$,
\begin{equation}
  -\1\leq \hat{A}_k\leq \1, \quad -\1\leq \hat{B}_m\leq \1, \quad [\hat{A}_k,\hat{B}_m]=0.
\end{equation}
It states that for any bipartite state $\rho$ for each of these operators there is a corresponding
real vector $\vec{x}_k$ (for $\hat{A}_k$) or $\vec{y}_m$ (for $\hat{B}_m$) such that $|\vec{x}_k|\leq 1$, $|\vec{y}_m|\leq 1$ and
\begin{equation} \label{tsi}
\langle \vec{x}_k,\vec{y}_m\rangle = \tr(\hat{A}_k\otimes\hat{B}_m\rho).
\end{equation}
This theorem also says that for any two sets of vectors $\{\vec{x}_k\}$ and $\{\vec{y}_m\}$ such that $|\vec{x}_k|\leq 1$ and
$|\vec{y}_m|\leq 1$ there exists a state and two sets of hermitian operators such that (\ref{tsi}) holds. Some of these operators can be projectors and other observables. For example, the vector version of the equation (\ref{hatA}) reads $\vec{x}_{\vec{a}}=\vec{x}_{\vec{a},1}-\vec{x}_{\vec{a},-1}$, where the vector $\vec{x}_{\vec{a}}$ corresponds to the observable and the other two to projectors. The lengths of the vectors corresponding to projectors is equal to the probability of the outcome associated with the given projector. In our case it is $\frac{1}{2}$ for all the vectors. This implies that the vector corresponding to the identity operator of Alice $\vec{x}_{\openone}$ is orthogonal to all the vectors corresponding to the projective measurements. Since Alice's outcomes are random she never uses the identity as a part of her strategy. What we need is that this operator can, in principle, be used. Obviously, if Alice and Bob both use the identity operators on their parts of the entangled state their results are always both 1 and therefore perfectly correlated, which means that $\vec{x}_{\openone}=\vec{y}_{\openone}$.

Therefore, we can write
\begin{equation}\begin{split}
  \langle B(\vec{a},\vec{b},A)\rangle
       &= \tr\Big(\big(\openone +A\hat{A}_{\vec{a}}\big)\otimes \hat{B}_{\vec{b},A}\rho\Big) \\
       &= \langle \vec{x}_{\openone} +A\vec{x}_{\vec{a}},\vec{y}_{\vec{b},A}\rangle          \\
       &=\langle \vec{x}_{\openone} +A\vec{x}_{\vec{a}},c(\vec{b},A)\vec{y}_{\openone}
                                               +c'(\vec{b},A)\vec{y}_{\perp\vec{b},A}\rangle \\
       &=c(\vec{b},A)+c'(\vec{b},A)\langle A\vec{x}_{\vec{a}}, \vec{y}_{\perp\vec{b},A}\rangle,
  \label{target}
\end{split}\end{equation}
where $c(\vec{b},A)=\langle \vec{y}_{\vec{b},A},\vec{y}_{\openone}  \rangle$, $c'(\vec{b},A)=\langle \vec{y}_{\vec{b},A},\vec{y}_{\perp\vec{b},A}  \rangle$ and $\vec{y}_{\perp\vec{b},A}$ is the projection of
$\vec{y}_{\vec{b},A}$ on the subspace orthogonal to $\vec{y}_{\openone}$.

It is now easy to see that the same value can be obtained with the
communication of a single hyperbit. Alice and Bob share a random bit
$A$ (with expectation value 0). Alice sends to Bob hyperbits $A \vec{x}_{\vec{a}}$ and Bob measures them using the vectors
$\frac{1}{N(\vec{b},A)}\vec{y}_{\perp\vec{b},A}$, where $N(\vec{b},A)$ is the norm of $\vec{y}_{\perp\vec{b},A}$. It is straightforward to calculate that Bob can obtain the value (\ref{target}) by the following postprocessing: First he discards his outcome with the probability $|c(\vec{b},A)|$ and outputs $sgn(c(\vec{b},A))$ instead. If he did not discard his outcome he flips it with probability $\frac{1}{2}\left(1-\frac{c'(\vec{b},A)N(\vec{b},A)}{1-|c(\vec{b},A)|} \right)$ and obtains (\ref{target}).

Proving conversely that communication of a single hyperbit can be simulated by shared entanglement and a single bit of classical
communication is quite straightforward. The most general strategy for players is for Alice to prepare the hyperbit in the state
$\vec{X}(\vec{a})$ and send it to Bob who will measure it with the vector $\vec{M}(\vec{b})$. His expectation value is
\begin{equation}
\langle B(\vec{a},\vec{b})\rangle=\langle \vec{X}(\vec{a}),\vec{M}(\vec{b})\rangle.
\end{equation}
Tsirelson's~\cite[Theorem 2.1]{Tsirelson:vectors} states that there exists a state and the collection of measurements such that
\begin{equation}
\langle \vec{X}(\vec{a}),\vec{M}(\vec{b})\rangle=\tr(\hat{A}_{\vec{a}}\otimes\hat{B}_{\vec{b}}\rho)=\langle AB\rangle,
\end{equation}
where $A$ is again Alice's outcome and $B$ is Bob's. It suffices now for Alice, after making the measurement, to transmit its outcome and Bob can, by multiplying the received message by his outcome, obtain the answer with exactly the same expectation value.

\section{Appendix B: Proof of Theorem 2}

The unbiasedness assumption means that
\begin{equation}
\sum_j f_{i,j}f_{i',j}=2^n \delta_{i,i'},
\end{equation}
i.e.~the matrix $F$ has orthogonal rows.
%
%
%
It also implies that not only the rows, but also the columns of $F$ are orthogonal:
\begin{equation} \label{ort}
\sum_i f_{i,j}f_{i,j'}=2^n \delta_{j,j'}.
\end{equation}

Let $\vec{h}_j$ be the hyperbit that Alice prepares when she gets input $j$.
The average hyperbit, that Bob receives when $I_i=\beta$ is
\begin{equation}
\vec{x}_{i,\beta}=\frac{1}{P(I_i=\beta)}\sum_j \frac{|f_{i,j} + \beta|}{2}p_j \vec{h}_j.
\end{equation}
If $\vec{i}$ is the vector that he measures with when he is interested in $I_i$, then
\begin{equation}\begin{split}
  P&(B_i|I_i=\beta) \\
   &= \frac{1}{2}\left(1+B_i\left\langle \vec{i},\frac{1}{P(I_i=\beta)}\sum_j \frac{|f_{i,j}+\beta|}{2}p_j
  \vec{h}_j\right\rangle \right).
  \label{probcon}
\end{split}\end{equation}
Then we have,
\begin{equation}\begin{split}
  E_i &= 2\bigl( P(B_i=1|I_i=1)P(I_i=1) \\
      &\phantom{==} +P(B_i=-1|I_i=-1)P(I_i=-1) \bigr) - 1  \\
      &= P(I_i=1)+\left\langle \vec{i},\sum_j \frac{|f_{i,j}+ 1|}{2}p_j \vec{h}_j\right\rangle \\
      &\phantom{==}
        +P(I_i=-1)-\left\langle \vec{i},\sum_j \frac{|f_{i,j}- 1|}{2}p_j \vec{h}_j\right\rangle - 1
\end{split}\end{equation}
Since $P(I_i=1)+P(I_i=-1)=1$, we get
\begin{equation}\begin{split}
  E_i &= \left\langle \vec{i},\sum_j \frac{|f_{i,j}+1|}{2}p_j \vec{h}_j\right\rangle
           -\left\langle \vec{i},\sum_j \frac{|f_{i,j}-1|}{2}p_j \vec{h}_j\right\rangle \\
      &= \left\langle \vec{i},\sum_j f_{i,j}p_j \vec{h}_j\right\rangle.
\end{split}\end{equation}
The optimal strategy for Bob is to choose his measurement such that $\vec{i}$ is parallel to $\vec{x}_i=\sum_j f_{i,j}p_j \vec{h}_j$. In
this case $E_i=|\vec{x}_i|$ or, equivalently,
\begin{equation}
E_i^2=\langle \vec{x}_i,\vec{x}_i \rangle.
\end{equation}
Let us calculate
\begin{equation}
\sum_i E_i^2=\sum_i\sum_{j,j'} f_{i,j}p_j f_{i,j'}p_{j'} \langle\vec{h}_j, \vec{h}_{j'}\rangle.
\end{equation}
Looking at eq.~(\ref{ort}), we have thus proved that
\begin{equation}
\sum_{i=1}^{2^n} E_i^2=2^n\sum_{j=1}^{2^n} p_j^2 |\vec{h}_j|^2.
\end{equation}

\section{Appendix C: Proof of Theorem 3}

We now prove, using hyperbits as a tool, that eq.~(\ref{ic}) holds in quantum mechanics
for pairwise independent, uniformly distributed bits $a_i$.
For every $i$,
\begin{equation}\begin{split}
  I(a_i:b_i) &= H(a_i)-H(a_i|b_i)   \\
             &= 1-P(b_i=1)H(a_i|b_i=1) \\
             &\phantom{=}
              -P(b_i=-1)H(a_i|b_i=-1)  \\
             &= P(b_i=1)\bigl( 1-h\bigl(P(a_i=1|b_i=1)\bigr)\bigr) \\
             &\phantom{=}
              +P(b_i=-1)\bigl( 1-h\bigl(P(a_i=-1|b_i=-1)\bigr)\bigr),
\end{split}\end{equation}
where $h(t) = -t\log t - (1-t)\log(1-t)$ is Shannon's binary entropy function.

Now, using the Taylor expansion
\begin{equation}\begin{split}
  1-h\left(\frac{1+x}{2} \right)
         &=    \sum_{n=1}^\infty \frac{1}{2n(2n-1)\ln(2)}x^{2n} \\
         &\leq \sum_{n=1}^\infty \frac{1}{2n(2n-1)\ln(2)}x^2
          =    x^2,
\end{split}\end{equation}
we obtain
\begin{equation}\begin{split}
  \label{i1}
  I(a_i:b_i) &\leq P(b_i=1)\left( 2P(a_i=1|b_i=1)-1\right)^2  \\
             &\phantom{:}
                   +P(b_i=-1)\left( 2P(a_i=-1|b_i=-1)-1\right)^2.\ \phantom{=}
\end{split}\end{equation}

From basic probability theory (Bayes' rule) we have
\begin{equation} \nonumber
P(a_i=\pm1|b_i=\pm1)= \frac{P(b_i=\pm1|a_i=\pm1)P(a_i=\pm1)}{P(b_i=\pm1)}.
\\
\end{equation}
Using the fact that sharing entanglement and communicating one classical bit is
in this scenario equivalent to communication of one hyperbit we can write
the probability using the formula (\ref{probcon}),
\begin{equation}\begin{split}
  P&(b_i=\beta|a_i=\alpha) \\
   &
    =\frac{1}{2}\left(1+\beta\left\langle\vec{i},\frac{1}{NP(a_i=\alpha)}\sum_j\frac{|f_{i,j}+\alpha|}{2} \vec{h}_j \right\rangle
    \right),\ \phantom{=}
\end{split}\end{equation}
and substituting $p_j=\frac{1}{N}$ we also have $P(a_i=\pm1)=\frac{1}{2}$ and
\begin{equation}
P(b_i=\beta)=\frac{1}{2}\left(1+\beta\left\langle\vec{i},\frac{1}{N}\sum_j \vec{h}_j \right\rangle \right).
\end{equation}
\begin{widetext}
Putting all this together we get
\begin{equation}\begin{split}
  \left( 2P(a_i=\alpha|b_i=\beta)-1\right)^2
     &= \left( 2\frac{P(b_i=\beta|a_i=\alpha)P(a_i=\alpha)}{P(b_i=\beta)}-1\right)^2
      = \left(\frac{\frac{1}{2}\left(1+\beta\left\langle\vec{i},\frac{2}{N}\sum_j\frac{|f_{i,j}+\alpha|}{2} \vec{h}_j \right\rangle
   \right)}{P(b_i=\beta)}-1\right)^2 \\
     &= \left(\frac{\frac{1}{2}\left(1+\beta\left\langle\vec{i},\frac{1}{N}\sum_j(|f_{i,j}+\alpha|) \vec{h}_j \right\rangle
   \right)-\frac{1}{2}\left(1+\beta\left\langle\vec{i},\frac{1}{N}\sum_j \vec{h}_j \right\rangle \right)}{P(b_i=\beta)}\right)^2 \\
     &= \frac{\left(\frac{1}{2}\left\langle\vec{i},\frac{1}{N}\sum_j f_{i,j} \vec{h}_j \right\rangle\right)^2 }{P(b_i=\beta)^2}.
\end{split}\end{equation}
\end{widetext}
Now we plug this into eq.~(\ref{i1}):
\begin{equation}\begin{split}
I(a_i:b_i) &\leq \frac{\left(\frac{1}{2}\left\langle\vec{i},\vec{x}_{i}
\right\rangle\right)^2\left(P(b_i=1)+P(b_i=-1)\right)}{P(b_i=1)P(b_i=-1)} \\
           &=    \frac{\frac{1}{4}\left(\left\langle\vec{i},\vec{x}_{i}
           \right\rangle\right)^2}{\frac{1}{4}\left(1-\left\langle\vec{i},\vec{x}_{avg} \right\rangle^2 \right)}.
\end{split}\end{equation}

By the Cauchy-Schwarz inequality,
$\left\langle\vec{i},\vec{x}_{i} \right\rangle^2 \leq \langle\vec{x}_i,\vec{x}_{i} \rangle$.
Furthermore, there can be no more than $2^n-1$ pairwise independent bits, so
\begin{equation}
\sum_{i=1}^N I(a_i:b_i)\leq \sum_{i=1}^{2^n-1} I(a_i:b_i).
\end{equation}
From the previous section we know that
\begin{equation}
\sum_{i=1}^{2^n-1} \langle\vec{x}_i,\vec{x}_{i} \rangle=1-\langle\vec{x}_{avg},\vec{x}_{avg} \rangle.
\end{equation}
Finally,
\begin{equation}\begin{split}
  \sum_{i=1}^N I(a_i:b_i)
         &\leq \sum_{i=1}^{2^n-1} I(a_i:b_i) \\
         &\leq \sum_{i=1}^{2^n-1}\frac{\left(\left\langle\vec{i},\vec{x}_{i}
         \right\rangle\right)^2}{\left(1-\left\langle\vec{i},\vec{x}_{avg} \right\rangle^2 \right)} \\
         &\leq \sum_{i=1}^{2^n-1}\frac{\left\langle\vec{x}_{i},\vec{x}_{i} \right\rangle}{1-\left\langle\vec{x}_{avg},\vec{x}_{avg}
         \right\rangle}
          = 1,
\end{split}\end{equation}
which concludes the proof.

\end{document}